\documentclass[a4paper,11pt]{article}
\usepackage{pos}
\usepackage{enumitem}
\usepackage{lineno}
\usepackage{wrapfig}

\title{Determination of the energy scale of cosmic ray measurements using the Auger Engineering Radio Array}
 \ShortTitle{Determination of the cosmic ray energy scale with AERA}

\author*[ab]{Tim Huege}

\affiliation[a]{Institute for Astroparticle Physics, Karlsruhe Institute of Technology, P.O. Box 3640, Karlsruhe, Germany}
\affiliation[b]{Astrophysical Institute, Vrije Universiteit Brussel, Pleinlaan 2, 1050 Brussel, Belgium}

\onbehalf{for the Pierre Auger Collaboration$^c$}
\affiliation[c]{Observatorio Pierre Auger, Av.\ San Mart{\'\i}n Norte 304, 5613 Malarg\"ue, Argentina\\
Full author list: {\rm\url{https://www.auger.org/archive/authors_icrc_2025.html}}}

\emailAdd{spokespersons@auger.org}

\abstract{The accurate determination of the absolute energy scale in cosmic ray 
measurements is both a challenging and fundamentally important task. We
present how measurements of radio pulses from extensive air 
showers with the Auger Engineering Radio Array, combined with per-event 
simulations of radio emission using the CoREAS extension of CORSIKA, allow us 
to determine the energy scale of cosmic rays between $3\cdot 10^{17}$\,eV and 
several $10^{18}$\,eV.
Our analysis accounts for many factors, each of which is controlled on the 5\% 
level or better. The absolute calibration of the antennas and the entire 
analog signal chain builds on a Galactic calibration in combination with a 
detailed understanding of the antenna-gain patterns. Additional key elements 
include compensation for temperature-dependent signal amplification, 
continuous detector health monitoring, an active veto for thunderstorm 
conditions, an unbiased event reconstruction, and per-event atmospheric 
modeling in the simulations. The analysis benefits from a high-statistics 
dataset of over 800 measured cosmic ray showers.
We describe our analysis method, perform multiple cross-checks, and evaluate 
systematic uncertainties. We find that absolute energies determined with AERA are 12\% higher
than those established with the Auger Fluorescence Detector, a result well in
agreement within systematic uncertainties and thus a strong independent confirmation
of the absolute energy scale of the Pierre Auger Observatory.}

\ConferenceLogo{PoS_ICRC2025_logo}

\FullConference{%
39th International Cosmic Ray Conference (ICRC2025)\\
15 -- 24 July, 2025\\
Geneva, Switzerland}

\begin{document}
\maketitle

%%%%%%%%%%%%%%%%%%%%%%%%%%%%%%%%%%%%%%%%%%%%%%%%%%%%%%%%%%%%%%%%%%%%%%%%%%%%%%%%%%%%%%%%%%

\section{Introduction}

A major challenge in the measurement of ultra-high-energy cosmic rays through air showers is the accurate determination of the absolute energy scale of the measured particles. At the Pierre Auger Observatory, the energy scale is determined with the Fluorescence Detector (FD). This approach avoids relying on simulations suffering from uncertainties in hadronic interaction models. In short, a well-calibrated measurement of the fluorescence light, directly proportional to the energy deposited by the air-shower particles in the atmosphere, is combined with precision atmospheric monitoring to determine the \emph{calorimetric energy} of individual air showers and then use this measurement to cross-calibrate the measurements with the water-Cherenkov detectors of the Surface Detector (SD) \cite{Dawson:2020bkp}. The FD energy scale underlying this work is the one underlying our most recent energy spectrum \cite{PierreAuger:2025hnw}, with the difference that here it is applied to the SD-750 array instead of the SD-1500 array. It reflects minor improvements over the most recent published SD-750 energy spectrum \cite{PierreAuger:2021hun}.

With the availability of radio antennas in the form of the \emph{Auger Engineering Radio Array (AERA)}, consisting of 153 autonomous detector stations covering an area of 17\,km$^2$ and measuring signals in the 30--80\,MHz band \cite{PierreAuger:2012ker}, an independent possibility arises to determine and validate the energy scale of the observatory. Fundamentally, the approach is similar to the one based on the FD, as also the total \emph{radiation energy} emitted by an air shower\footnote{The total energy radiated in the form of radio waves, in the case of AERA in the 30--80\,MHz band.}, after applying well-known density corrections, is a calorimetric measurement of the energy in the electromagnetic cascade \cite{Glaser:2016qso}. A well-calibrated radio detector and knowledge of the expected \emph{radiation energy} for a given \emph{electromagnetic energy}, either from first-principles microscopic Monte-Carlo simulations or from laboratory measurements, can then provide an independent measure of the energy scale of the observatory \cite{PierreAuger:2016vya}.
In the following, we describe how we establish the energy scale of cosmic-ray measurements from radio signals in the energy range of $3 \cdot 10^{17}$ to several $10^{18}$\,eV with a high-quality data set of air showers measured with AERA and simulated with the CoREAS~\cite{Huege:2013vt} code.

%%%%%%%%%%%%%%%%%%%%%%%%%%%%%%%%%%%%%%%%%%%%%%%%%%%%%%%%%%%%%%%%%%%%%%%%%%%%%%%%%%%%%%%%%%

\section{Determination of the energy scale with AERA}

\begin{figure}
	\centering
	\includegraphics[width=0.68\textwidth]{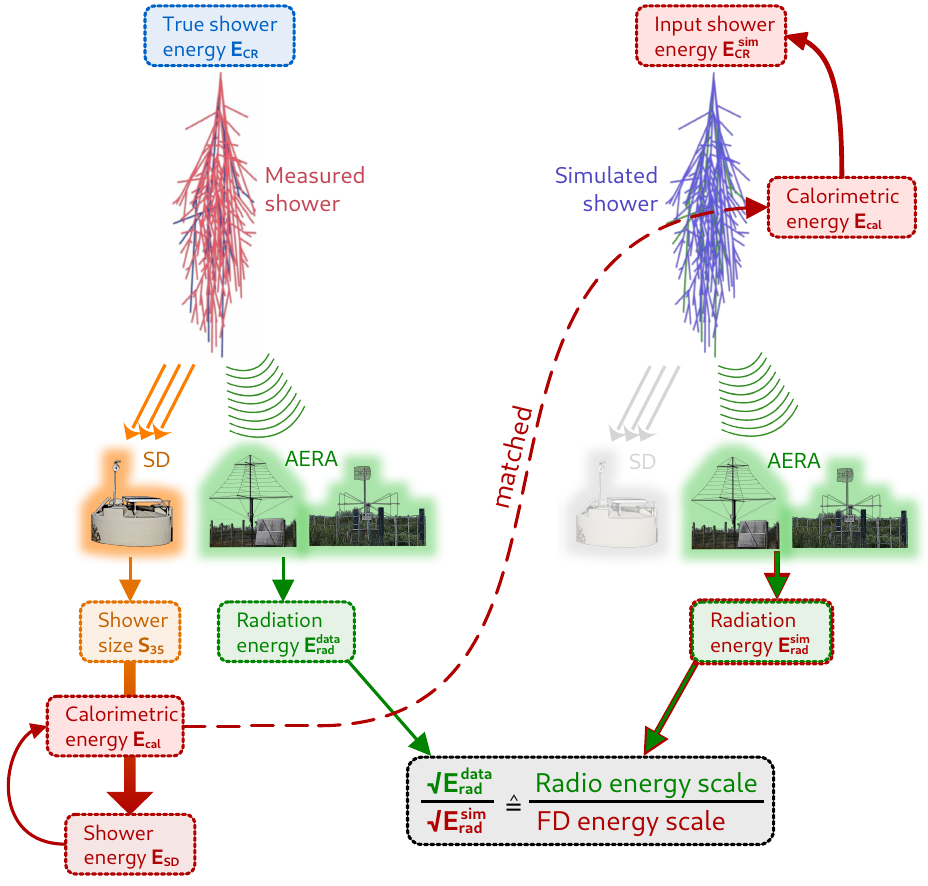}
	\caption{Logic for event-by-event comparison of measured and simulated air showers, determining the agreement between the FD and radio energy scales. Observables depicted in red are on the FD energy scale. The orange and green colors represent SD and radio measurements, respectively.}
	\label{fig:8_analysis_logic}
\end{figure}

The logic of the analysis is shown in Fig.\ \ref{fig:8_analysis_logic}. For an air shower measured with the SD, we reconstruct the \emph{shower energy} $E_\mathrm{SD}$, the absolute scale of which has previously been established with the FD. This energy, also called \emph{cosmic-ray energy}, estimates the total energy in the air shower, including contributions which are visible neither to the FD nor to radio antennas. The so-called \emph{invisible energy} has previously been determined in a data-driven way as a function of \emph{cosmic-ray energy} \cite{PierreAuger:2019dhr}. After subtracting the \emph{invisible energy} from the \emph{shower energy}, we have an estimate of the \emph{calorimetric energy}, i.e., the total energy deposited in the atmosphere by the electromagnetic cascade, on the energy scale set by the FD. Using the air-shower geometry and this \emph{calorimetric energy} estimate from the SD, we then set up a CORSIKA \cite{HeckKnappCapdevielle1998} simulation of the corresponding air shower, along with radio-emission simulations with the CoREAS \cite{Huege:2013vt} extension, such that the resulting simulated shower will have the target \emph{calorimetric energy}.

In a final step, we reconstruct the \emph{radiation energy} measured with AERA \cite{PierreAuger:2015hbf} for both the measured shower (which is on the \emph{radio energy scale}) as well as for the simulated shower (which is on the \emph{FD energy scale}, as the input energy for the simulation is based ultimately on the \emph{FD energy scale}). The ratio of the reconstructed \emph{radiation energies} reconstructed from the measurements and simulations then determines the ratio between the \emph{radio energy scale} and the \emph{FD energy scale}.

While the logic of this analysis is seemingly simple, the challenge lies in the fact that systematic uncertainties on all involved detectors and their corresponding event reconstructions need to be understood on a 5\% level or better to provide a result with the desired accuracy. In terms of the absolute calibration of AERA, this requires in particular knowledge of the frequency-dependent antenna-gain patterns of the used LPDA and butterfly antennas \cite{PierreAuger:2012ker} (known to within 5\%) and the absolute calibration of the radio measurements from a comparison of the Galactic radio emission (known at a level of 6\%) with the measured radio emission over full local sidereal days \cite{Busken:2022mub,Santos:2024ldc}. The AERA event reconstruction \cite{PierreAuger:2015hbf} of the \emph{radiation energy} has previously been demonstrated to be bias-free irrespective of the type of primary particle. The signal processing with the Offline analysis framework \cite{PierreAuger:2011btp} is carried out in the exact same way on simulations and data, so that any minor biases possibly arising from analysis steps such as bandpass filtering or signal cleaning affect data and simulations in the same way and thus cancel out in the final ratio. 

%%%%%%%%%%%%%%%%%%%%%%%%%%%%%%%%%%%%%%%%%%%%%%%%%%%%%%%%%%%%%%%%%%%%%%%%%%%%%%%%%%%%%%%%%%

\section{Event selection and simulation production}

We begin with a data set of 3,245 air showers measured in the period from 2014 to 2020 with the 750\,m array of the SD, passing standard SD quality cuts, having their cores contained within the geometrical area of the AERA stations capable of accepting an SD trigger, and having reconstructed SD energies of at least $3 \cdot 10^{17}$\,eV, thus ensuring full detection efficiency of the SD. Exclusion of confirmed thunderstorm periods \cite{Gottowik2021} or periods with no data on thunderstorm activity reduces the data set to 3,015 air showers. For these 3,015 events, we perform a reconstruction of the \emph{radiation energy} $E_{\mathrm{rad}}$ from the data of the AERA stations that are able to receive an external trigger from the SD and were fully functional at the time of measurement. The main criterion that 5~AERA stations had a signal amplitude a factor of $\sqrt{10}$ above the noise-RMS reduces the data set to 1,096 air showers.  Requiring some quality cuts on reconstructed uncertainties results in a data set of 902 air showers with a valid $E_{\mathrm{rad}}$ reconstruction. Applying further cuts to ensure high data quality as listed in the upper half of Tab.\ \ref{tab:8_quality_event_selection} results in a data set of 844 air showers on which the remainder of the analysis builds.

%\begin{table*}[h!]
%	\caption{Event cuts for the pre-selection of the hybrid dataset in the preparation of matched simulations. The survival fractions always relate to the previous selection step.}
%	\label{tab:8_event_preselection}
%	\centering
%	\begin{tabular}{l r r}
%	\hline\hline
%	\textbf{Cut or requirement} & $N_\mathrm{events}$ & Survival fraction \\ \hline
%
%        \textbf{SD cuts + cores confined in AERA}                       &  &  \\
%        \quad $E_\mathrm{SD} > 3 \times 10^{17}$\,eV                           & 3245 & - \\
%        \quad Thunderstorm veto                                         & 3015 & 92.9 \% \\
%        \textbf{Base radio cuts}                                        &  &  \\
%        \quad $N_\mathrm{stations, signal} \geq 5$                      & 1096 & 36.4 \% \\
%        \quad $E_\mathrm{rad}$ reconstructed                            & 1062 & 96.9 \% \\
%        \quad $\sigma_{E_\mathrm{rad}}$ reconstructed                   & 902  & 84.9 \% \\
%        \quad Radio core reconstructed                                  & 902  & 100.0 \% \\
%        
%        \hline
%        \hline
%	\end{tabular}
%\end{table*}

For each of these 844 air showers we prepare one CoREAS simulation each with a proton and an iron primary according to the logic described above. The chosen interaction models are Sybill 2.3d and UrQMD. We note, however, that our result does not depend on the choice of hadronic interaction model as the radio emission is purely sensitive to the electromagnetic cascade of the air shower and we subtract the \emph{invisible energy} of the chosen hadronic interacion model in a self-consistent way. For the  simulations, we adopt a per-event atmospheric density and refractivity gradient \cite{Mitra:2020mza} derived from the Global Data Assimilation System (GDAS). Furthermore, particle thinning at the $10^{-6}$ level with optimized weight limitation and a fine-grained treatment of multiple scattering (STEPFC parameter of 0.05) \cite{Gottowik:2017wio} ensure absolute signal predictions with the highest-possible accuracy. The simulated radio signals are then fed through a detailed detector simulation with the Offline analysis framework \cite{PierreAuger:2011btp}, including the addition of measured background noise taken at a given antenna station within 10 minutes of the measured event, before being reconstructed in the exact same way as the data.

\begin{figure}
	\centering
	\includegraphics[width=0.44\textwidth]{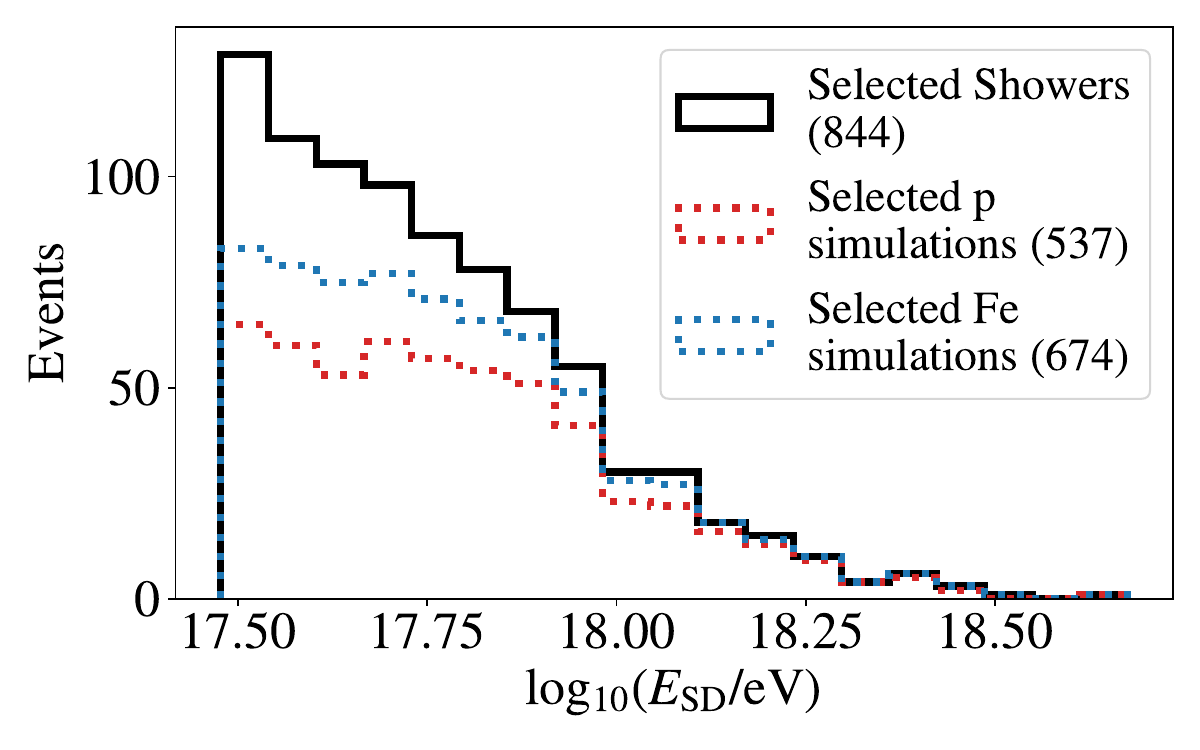}
    \includegraphics[width=0.27\textwidth]{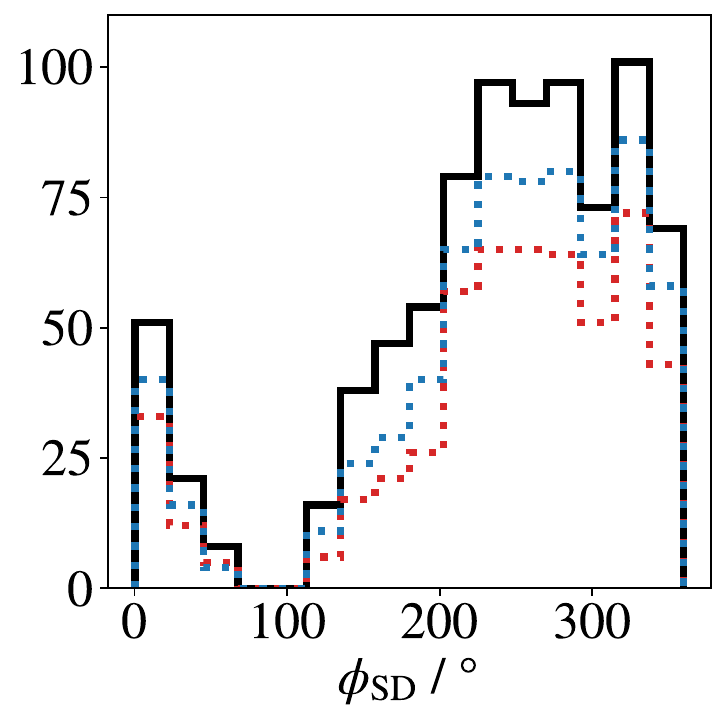}
    \includegraphics[width=0.27\textwidth]{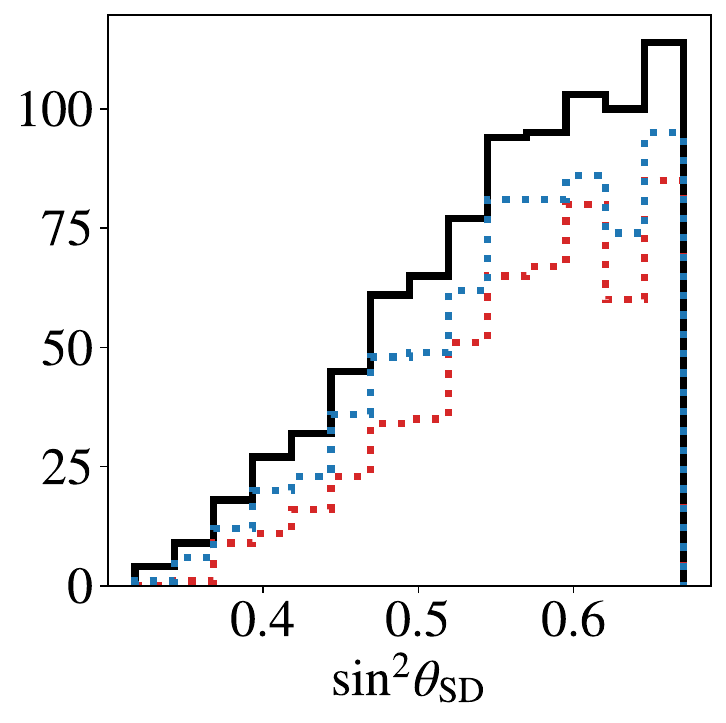}
	\caption{Left: Distribution of the logarithmic SD-reconstructed energies of the quality event set of 844 air showers and the proton and iron simulation sets after reconstruction and quality cuts. Middle and Right: Distributions of the azimuth angle, counter-clockwise from geographic East, and the zenith angle, as reconstructed with the SD for the same data and simulation sets.}
	\label{fig:8_Eventset_energies}
\end{figure}

As expected, not all simulations pass the $E_{\mathrm{rad}}$ reconstruction, since AERA is not fully efficient at the lower end of this energy range. In particular, deep low-energy showers might not illuminate the required 5~AERA stations with sufficient signal strength. The survival probability for iron-induced showers is higher than that of proton showers. The AERA data set will thus no longer be unbiased. As the radio emission is only associated with the \emph{electromagnetic energy}, this is not a critical problem, in a similar way that the chosen hadronic interaction model does not influence the result of the analysis. However, this potential shift towards a heavier mass composition could introduce second-order effects on the mean of the reconstructed SD energies, which we have estimated from simulations to be less than 2.4\%, and which we thus account for with a corresponding systematic uncertainty. The same additional quality cuts previously applied to the measured data and listed in Tab.\ \ref{tab:8_quality_event_selection} reduce the total number of usable simulations to 537 and 674 for proton- and iron-induced air showers, respectively. The distributions of $E_\mathrm{SD}$, arrival azimuth angle and arrival zenith angle for the 844 selected events as well as the surviving simulations for proton and iron primaries are shown in Fig.\ \ref{fig:8_Eventset_energies}.

We note that during the complete development of the analysis (concerning the choice of cuts, optimization of the event reconstruction, etc.), we applied an unknown, global ``blinding factor'' to the measured radiation energies, as to not unconsciously influence the result in a ``favorable'' direction. This blinding factor was only lifted after the analysis was completely defined.

\begin{table*}[h!]
	\caption{Quality selection cuts for the event set. Survival fractions relate to the previous selection step. $\chi^2 / \mathrm{ndf}$ refers to the quality of the LDF fit, $\alpha$ is the geomagnetic angle.}
	\label{tab:8_quality_event_selection}
	\centering
    \small
	\begin{tabular}{l r r}
	\hline\hline
	\textbf{Cut or requirement} & $N_\mathrm{events}$ & Survival fraction \\ \hline

        After SD cuts \& base cuts (see text)  & 902 &  \\
        \textbf{Additional radio cuts on data}                          &      &  \\
        \quad No saturated stations within 500\,m                       & 873  & 96.8\% \\
        \quad $\chi^2 / N_\mathrm{df} < 10$                             & 854  & 97.8\% \\
    \vspace{0.1cm}
        \quad $\alpha > 20^{\circ}$                                     & 844  & 98.8\% \\
%                                                                        &      &  \\
%        Simulation available p / Fe                                     & 792 / 817 & 93.8 \% / 96.8 \% \\
        \textbf{Radio cuts on simulations}                              &      &  \\
        \quad $N_\mathrm{stations, signal} \geq 5$                      & 549 / 684 &  65.0\% /  81.0\% \\
        \quad $E_\mathrm{rad}$ reconstructed                            & 542 / 678 &  98.7\% /  99.1\% \\
        \quad $\sigma_{E_\mathrm{rad}}$ reconstructed                   & 538 / 677 &  99.3\% /  99.9\% \\
        \quad Radio core reconstructed                                  & 538 / 676 & 100.0\% /  99.9\% \\
        \quad No saturated stations within 500\,m                       & 538 / 676 & 100.0\% / 100.0\% \\
        \quad $\chi^2 / \mathrm{ndf} < 10$                              & 537 / 676 &  99.8\% / 100.0\% \\
        \quad $\alpha > 20^{\circ}$                                     & 537 / 674 & 100.0\% /  99.7\% \\
        \hline
        \hline
	\end{tabular}
\end{table*}

%%%%%%%%%%%%%%%%%%%%%%%%%%%%%%%%%%%%%%%%%%%%%%%%%%%%%%%%%%%%%%%%%%%%%%%%%%%%%%%%%%%%%%%%%%

\section{Results on the energy scale}

%\begin{figure}
%    \centering
%    \includegraphics[width=0.35\textwidth]{figures/energy_comp_logRatio_data_v31r15p3_sim_v33r0_5nS_0zenith_3e+17minE_p_corrected.pdf}
%    \includegraphics[width=0.35\textwidth]{figures/energy_comp_logRatio_data_v31r15p3_sim_v33r0_5nS_0zenith_3e+17minE_Fe_corrected.pdf}
%\caption{Distribution of the logarithmic square root ratios of the reconstructed radiation energies from measured events and the corresponding simulations. \textit{Left:} results for proton simulations, \textit{Right:} results for iron simulations.}
%    \label{fig:8_Result}
%\end{figure}

\begin{figure}[htbp]
    \centering
    \begin{minipage}[b]{0.48\textwidth}
        \centering
        \includegraphics[width=\textwidth]{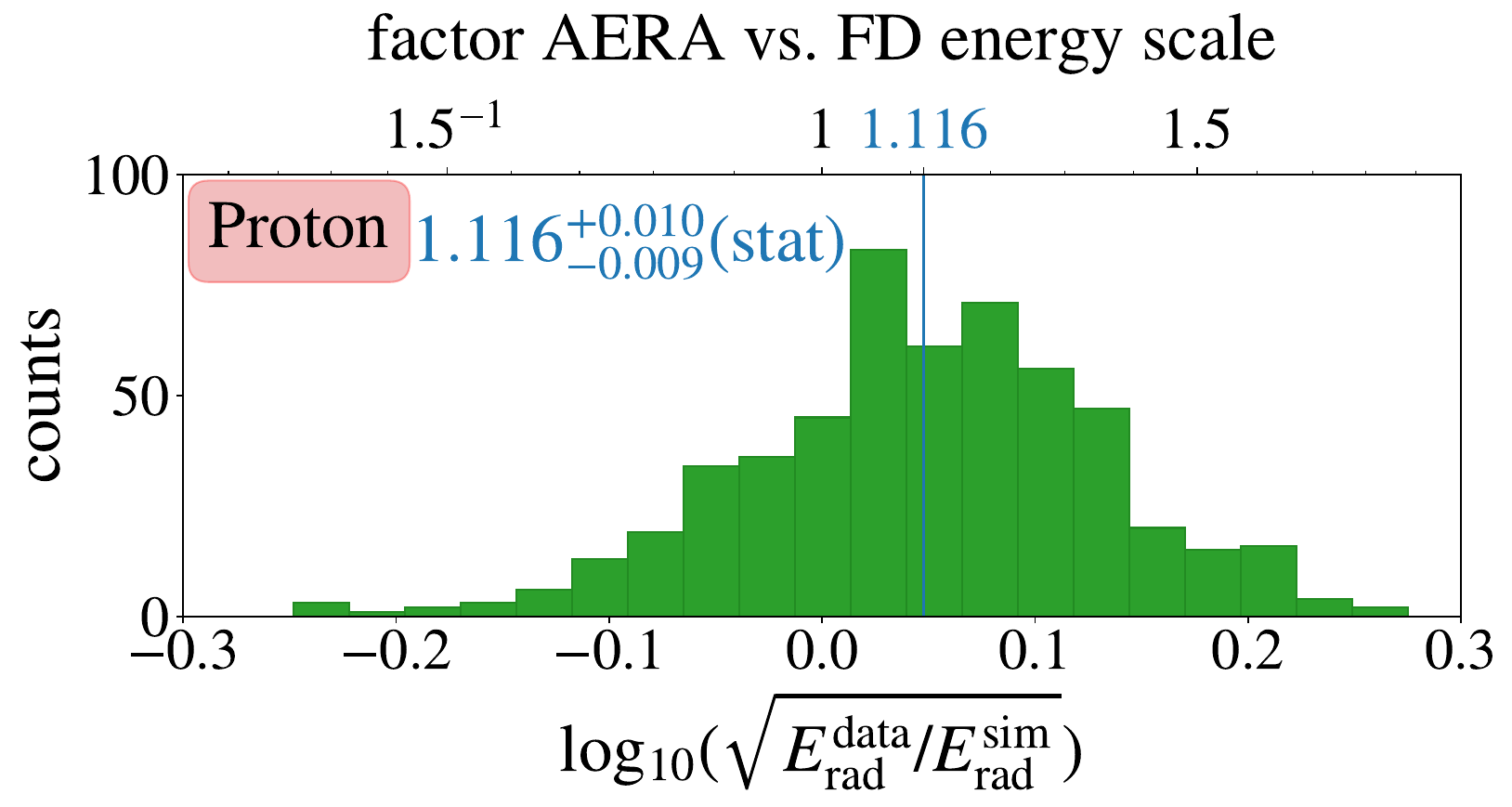}
    \end{minipage}
    \hfill
    \begin{minipage}[b]{0.48\textwidth}
        \centering
        \includegraphics[width=\textwidth]{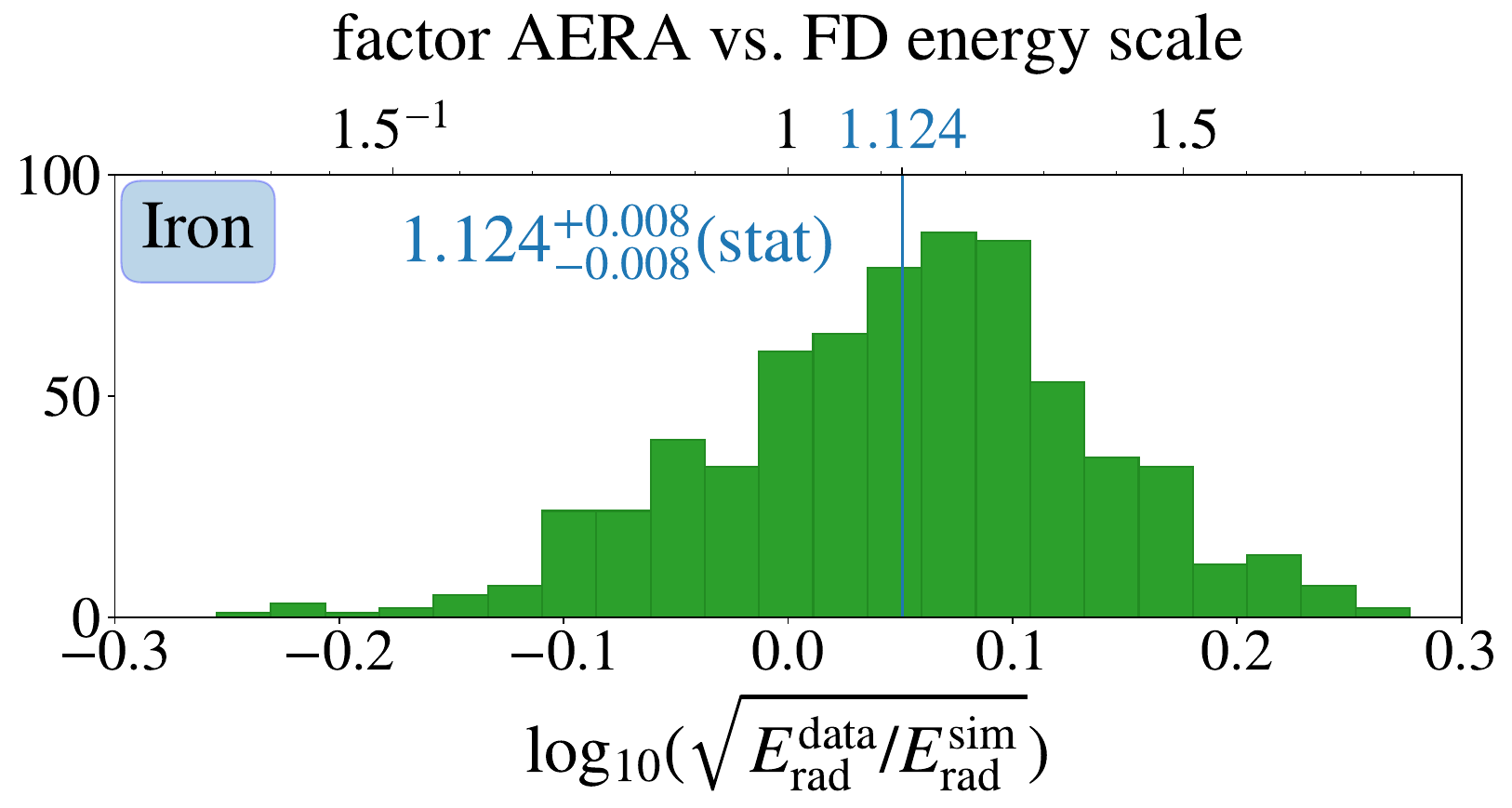}
    \end{minipage}
    \caption{Distribution of the logarithm of the square root ratios of the reconstructed radiation energies for measured events with the two simulation sets. \textit{Left:} result for proton simulations, \textit{Right:} result for iron simulations. The resulting mean scale factors are indicated on the top axes.}
    \label{fig:8_Result}
\end{figure}

In Fig.\ \ref{fig:8_Result}, we present histograms of the ratios of \emph{radiation energies} for measured and simulated events, once for the proton and once for the iron simulations. The square root of the ratio is taken because the \emph{radiation energy} scales with the square of the \emph{electromagnetic energy} due to the coherent nature of the radio emission. We bin in $\log_{10}$ of the ratios rather than simple ratios, thereby ensuring symmetry between ratios smaller and larger than unity. The linear scale factors resulting from the means of the shown distributions for proton and iron simulations are indicated at the top axes of the diagrams. The two scale factor are compatible within statistical uncertainties -- as expected, there is no systematic difference induced by the type of primary particle. Averaging the two mean values and afterwards compensating for the $\log_{10}$ operation yields that
\begin{equation} \label{eq:8_result_energy_scales_statonly}
    \frac{\mathrm{Radio\ energy\ scale}}{\mathrm{FD\ energy\ scale}} = 1.120\,^{+0.009}_{-0.008}\,\mathrm{(stat)},
\end{equation}
where the quoted $\sim 0.8$\% uncertainty refers to statistical uncertainties only. This result signifies that the energy scale determined from radio data is 12\% higher than the FD energy scale. In other words: A $10^{18}$\,eV air shower measured on the FD energy scale has an energy of $1.12 \cdot 10^{18}$\,eV on the energy scale determined from radio data.

\begin{figure}[htbp]
    \centering
    \begin{minipage}[b]{0.48\textwidth}
        \centering
        \includegraphics[width=\textwidth]{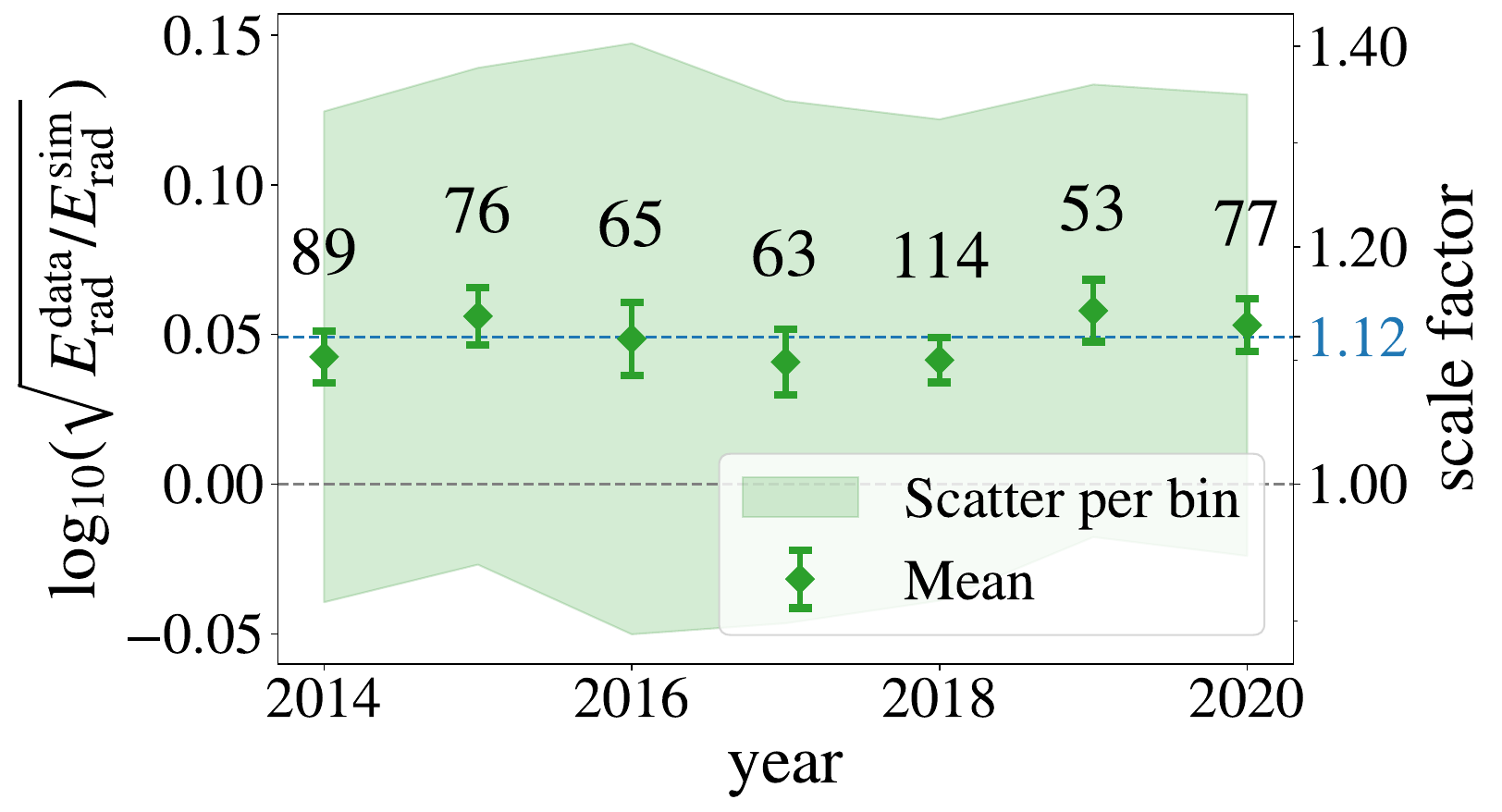}
    \end{minipage}
    \hfill
    \begin{minipage}[b]{0.48\textwidth}
        \centering
        \includegraphics[width=\textwidth]{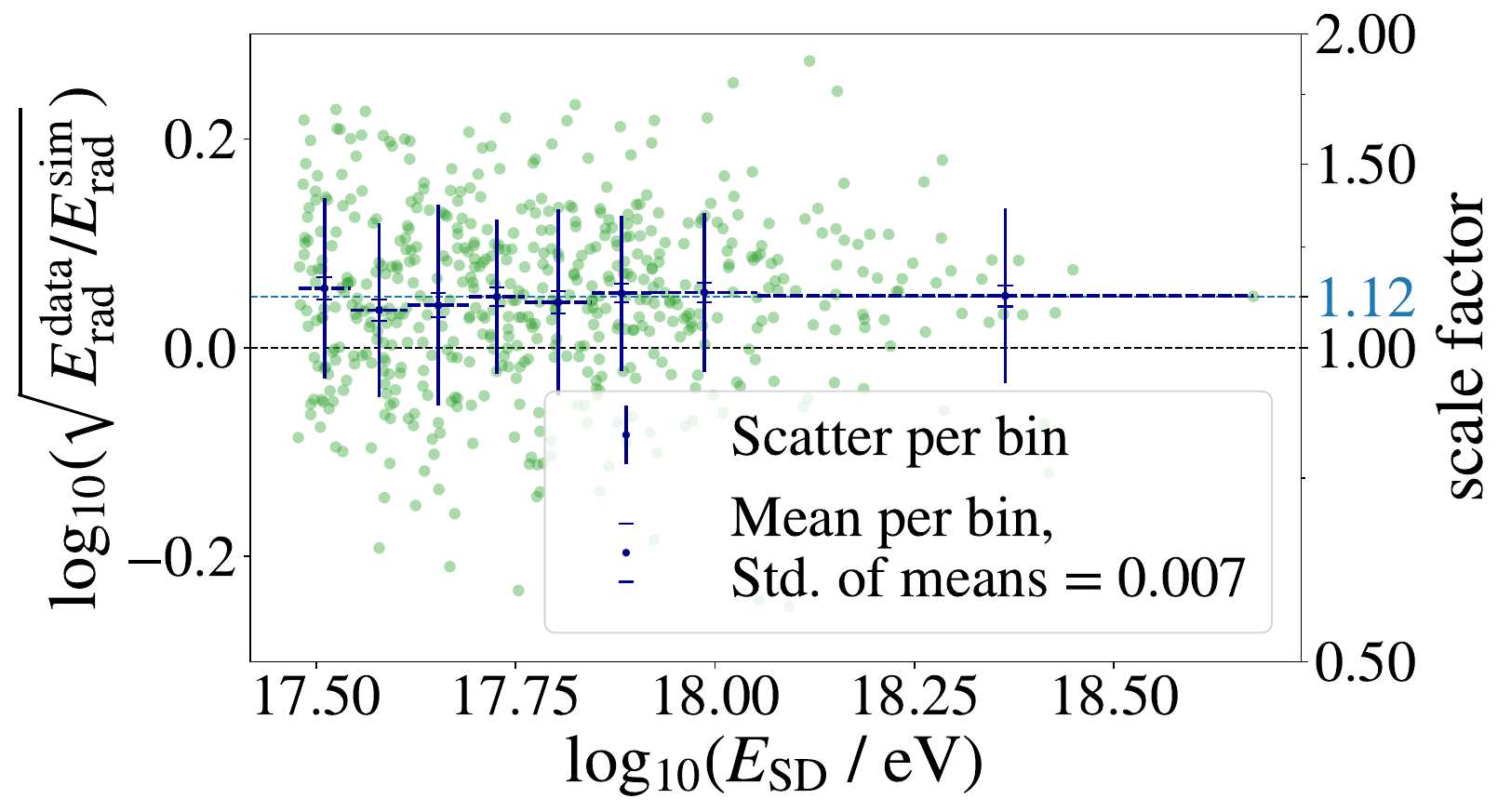}
    \end{minipage}
    \caption{\textit{Left:} Evolution of the mean of the logarithmic square root ratio of radiation energies across the years covered by the event set, for proton simulations. \textit{Right:} Dependence of the mean logarithmic square root ratio of radiation energies on the SD-reconstructed shower energy.}
    \label{fig:8_Syschecks}
\end{figure}

We have stratified our data in many different ways to check for potential correlations with arrival direction, core position, ambient temperature, year of data taking, number of signal stations and SD-reconstructed shower energy. None of the tests showed any significant correlations. In Fig.\ \ref{fig:8_Syschecks}, we show the $\log_{10}$ of the square root ratios as a function of time (left) and $E_\mathrm{SD}$ (right). The fact that there is no time-evolution confirms that the radio measurement is not suffering from detector ageing \cite{Santos:2024ldc}. The fact that the scale factor remains constant over a decade in energy means that there is no slope between the FD and radio energy scales, another important and non-trivial result.

%\begin{figure}
    %\centering
	%\includegraphics[width=1.2\textwidth]{figures/energy_longterm_data_v31r15p3_sim_v33r0_5nS_0zenith_3e+17minE_p_corrected.pdf}
    %\includegraphics[width=1.2\textwidth]{figures/energy_dep_ESD_data_v31r15p3_sim_v33r0_5nS_0zenith_3e+17minE_p_corrected.pdf}
%\caption{\textit{Left:} Evolution of the mean and median square root ratio of radiation energies across the years covered by the event set, for proton simulations. The uncertainty on the median is calculated as the uncertainty on the mean multiplied by 1.253. \textit{Right:} Dependence of the square root ratio of radiation energies on the SD-reconstructed shower energy.}
    %\label{fig:8_Syschecks}
%\end{figure}

%\begin{equation} \label{eq:8_result}
%    \begin{aligned}
%        & \mathrm{\textbf{Proton:}\quad}        &10^{0.0478} &= 1.116 \\
%        & \mathrm{\quad Stat.\ unc.\ up:\quad}   &10^{0.0478 + 0.0036} &= 1.126\\
%        & \mathrm{\quad Stat.\ unc.\ down:\quad}   &10^{0.0478 - 0.0036} &= 1.107\\
%        & & & \\
%        & \mathrm{\textbf{Iron:}\quad}          & 10^{0.0507} &= 1.124 \\
%        & \mathrm{\quad Stat.\ unc.\ up:\quad}   &10^{0.0507 + 0.0032} &= 1.132\\
%        & \mathrm{\quad Stat.\ unc.\ down:\quad}   &10^{0.0507 - 0.0032} &= 1.116
%    \end{aligned}
%\end{equation}

%%%%%%%%%%%%%%%%%%%%%%%%%%%%%%%%%%%%%%%%%%%%%%%%%%%%%%%%%%%%%%%%%%%%%%%%%%%%%%%%%%%%%%%%%%

\section{Systematic uncertainties}

We have performed an in-depth investigation of systematic uncertainties affecting the result, listed in Tab.\ \ref{tab:8_uncertainties_summary}. Of the experimental uncertainties, the most important ones are the accuracy of the Galactic calibration arising from performing the calibration with a total of 7 available sky models, interpreting the spread of the resulting calibration constants as a systematic uncertainty amounting to 6.1\%. Uncertainties still remaining in our understanding of the antenna response pattern, in particular for the butterfly antennas, add an additional 5\%. Drone-based calibration campaigns are ongoing and have the potential to lower this uncertainty in the future. The systematic uncertainty resulting from a potential bias of the mass composition towards heavier elements due to inefficiency of AERA in detecting deep showers is conservatively estimated to 2.4\%. All other investigated uncertainties are negligible in comparison.

Of the uncertainties related to predicting the absolute strength of the radio emission, the dominant one with 5.0\% is the ``radio emission yield'', i.e., the relation between the absolute radio-emission strength and the energy in the underlying electromagnetic cascade. This value results from in-depth comparisons of independent implementations of both the simulations of extensive air showers and their associated radio emission with first-principle classical electrodynamics within CORSIKA 7, ZHAireS, CORSIKA 8 \cite{Gottowik:2017wio,Alameddine:2024cyd} and with the ``endpoints'' and ZHS formalisms \cite{Alameddine:2024cyd}. In terms of experimental verification, the only  measurement available so far is that of the SLAC T-510 experiment, which showed agreement between measured and simulated signals on the level of 5\% with a systematic uncertainty of $\sim10$\% \cite{Bechtol:2021tyd}. Again, other systematic uncertainties related to the signal prediction are negligible.

If one accepts the spread on the radio-emission between different microscopic simulations as the systematic uncertainty on the ``radio emission yield'', the total systematic uncertainty on the energy scale determined from AERA amounts to 10.2\%. Taking a more conservative approach based on the SLAC T-510 measurement results in a total systematic uncertainty of 13.4\%. The systematic uncertainty of the FD energy scale amounts to 14\%. Statistical uncertainties amount to 0.8\% and are thus negligible in comparison with systematic uncertainties. Assuming the uncertainties of the FD and AERA-determined energy scales to be uncorrelated and thus summing them in quadrature, for the case of the simulations-based radio-emission yield uncertainty, thus results in

\begin{equation} \label{eq:8_result_energy_scales}
    \frac{\mathrm{Radio\ energy\ scale}}{\mathrm{FD\ energy\ scale}} = 1.120 \pm{0.173}\,\mathrm{(sys)}\,  ^{+0.009}_{-0.008}\,\mathrm{(stat)}.
\end{equation}
The energy scales determined from the FD and AERA are thus well in agreement within systematic uncertainties.

\begin{table*}[h!]
	\caption{Summary of uncertainties on the radio energy scale.}
	\label{tab:8_uncertainties_summary}
	\centering
    \small
	\begin{tabular}{l c}
    \hline\hline
    \textbf{Source of uncertainty} & \textbf{Size} \\ \hline
    
    \textbf{Experimental uncertainties}                         & \textbf{8.8 \%} \\
    \quad Galactic calibration                                  & 6.1 \% \\
    \quad Antenna response pattern                              & 5 \% \\
    \quad LDF model                                             & 0.5 \% \\
    \quad Atmosphere                                            & <1.25 \% \\
    \quad Ground conditions                                     & <1.8 \% \\
    \quad Quiet sun                                             & 0.5 \% \\
    \quad Simulation primary bias                               & 1.9 \% \\
    % \quad \sout{Signal loss during reconstruction} ?            & \sout{0.14 \%} ? \\ 
    \vspace{0.2cm}
    \quad Mass composition bias w.r.t.~Auger mix                  & 2.4 \% \\
    %        \quad Biases on $\theta$, $X_\mathrm{max}$, $n_\mathrm{signal~stations}$ & $\sim$1\% \\
    
    \textbf{Theoretical uncertainties}                          & \textbf{5.1 \% / $\sim$10.1 \%} \\
    \quad Radio emission yield                                  & 5.0 \% / $\sim$10 \% \\
    \quad Choice of hadronic interaction model                  & 0.13 \% \\
    \quad Thinning                                              & <0.15 \% \\
    \vspace{0.2cm}
    \quad Energy thresholds of shower particles                 & <0.5 \% \\ 
    
    \textbf{Stat.~unc.~of energy scale comparison}       & \textbf{0.8 \%} \\ 
    
    \hline
    \textbf{Total absolute scale uncertainty}                   & \textbf{Stat.:} \textbf{0.8 \%}  ~ \rule[-2pt]{1pt}{10pt} \hspace{-5pt} \rule[-2pt]{1pt}{10pt} ~ \textbf{Syst.:} \textbf{10.2 \% / $\sim$13.4 \%} \\
    
    \hline
    \hline
  \end{tabular}
\end{table*}

%%%%%%%%%%%%%%%%%%%%%%%%%%%%%%%%%%%%%%%%%%%%%%%%%%%%%%%%%%%%%%%%%%%%%%%%%%%%%%%%%%%%%%%%%%

\section{Conclusions}

We have used a high-quality data set of 844 air showers measured with AERA in the energy range from $3 \cdot 10^{17}$\,eV to several $10^{18}$\,eV in comparison with simulations of the same air showers using the CoREAS code to independently determine the absolute energy scale of cosmic-ray measurements at the Pierre Auger Observatory using radio signals. Particular care has been taken in crafting an analysis that does not suffer from biases and in which all individual systematic uncertainties are under control at the 5\% level. The analysis was developed using a blinded data set to avoid subconscious influence on the analysis and the choice of selection cuts. The end result is that the energy scale determined from AERA is 12\% higher than the one from the FD, a constant factor over the investigated energy range and the period from 2014 to 2020. The systematic uncertainty on the energy scale determined from radio data amounts to 10.2\% when building on the ``radio emission yield'' uncertainty estimated from simulations, or 13.4\% when relying on the measurement uncertainty reported by the SLAC T-510 experiment. The energy scale derived from AERA is well in agreement with the Auger energy scale determined from measurements with the Fluorescence Detector, a strong and independent confirmation of the FD energy scale within systematic uncertainties. In the future, we will use the Auger Radio Detector covering the full 3,000\,km$^2$ of the observatory to also independently determine the energy scale at the highest energies from radio measurements.

%%%%%%%%%%%%%%%%%%%%%%%%%%%%%%%%%%%%%%%%%%%%%%%%%%%%%%%%%%%%%%%%%%%%%%%%%%%%%%%%%%%%%%%%%%

\let\oldbibliography\thebibliography
\renewcommand{\thebibliography}[1]{%
  \oldbibliography{#1}%
  \setlength{\itemsep}{1pt}%
}

%{\footnotesize
%\bibliographystyle{JHEPnotitle}
%\bibliography{references}
%}

\providecommand{\href}[2]{#2}\begingroup\raggedright\endgroup

%% Full authors list (ONLY FOR COLLABORATIONS)
\clearpage
\section*{The Pierre Auger Collaboration}
{\footnotesize\setlength{\baselineskip}{10pt}
\noindent
\begin{wrapfigure}[11]{l}{0.12\linewidth}
\vspace{-4pt}
\includegraphics[width=0.98\linewidth]{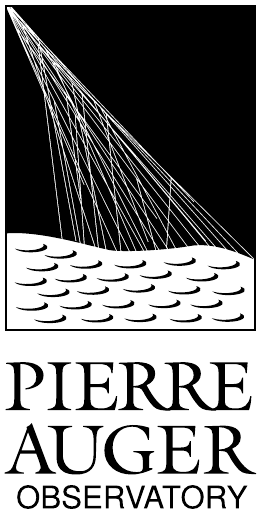}
\end{wrapfigure}
\begin{sloppypar}\noindent
% created on 2025-06-06
A.~Abdul Halim$^{13}$,
P.~Abreu$^{70}$,
M.~Aglietta$^{53,51}$,
I.~Allekotte$^{1}$,
K.~Almeida Cheminant$^{78,77}$,
A.~Almela$^{7,12}$,
R.~Aloisio$^{44,45}$,
J.~Alvarez-Mu\~niz$^{76}$,
A.~Ambrosone$^{44}$,
J.~Ammerman Yebra$^{76}$
% ...

\end{sloppypar}
\begin{center}
\end{center}

\vspace{1ex}
% created on 2025-06-06
% needs \usepackage{enumitem}
\begin{description}[labelsep=0.2em,align=right,labelwidth=0.7em,labelindent=0em,leftmargin=2em,noitemsep,before={\renewcommand\makelabel[1]{##1 }}]
\item[$^{1}$] Centro At\'omico Bariloche and Instituto Balseiro (CNEA-UNCuyo-CONICET), San Carlos de Bariloche, Argentina
\item[$^{2}$] Departamento de F\'\i{}sica and Departamento de Ciencias de la Atm\'osfera y los Oc\'eanos, FCEyN, Universidad de Buenos Aires and CONICET, Buenos Aires, Argentina
\item[$^{3}$] IFLP, Universidad Nacional de La Plata and CONICET, La Plata, Argentina
\item[$^{4}$] Instituto de Astronom\'\i{}a y F\'\i{}sica del Espacio (IAFE, CONICET-UBA), Buenos Aires, Argentina
\item[$^{5}$] Instituto de F\'\i{}sica de Rosario (IFIR) -- CONICET/U.N.R.\ and Facultad de Ciencias Bioqu\'\i{}micas y Farmac\'euticas U.N.R., Rosario, Argentina
\item[$^{6}$] Instituto de Tecnolog\'\i{}as en Detecci\'on y Astropart\'\i{}culas (CNEA, CONICET, UNSAM), and Universidad Tecnol\'ogica Nacional -- Facultad Regional Mendoza (CONICET/CNEA), Mendoza, Argentina
\item[$^{7}$] Instituto de Tecnolog\'\i{}as en Detecci\'on y Astropart\'\i{}culas (CNEA, CONICET, UNSAM), Buenos Aires, Argentina
\item[$^{8}$] International Center of Advanced Studies and Instituto de Ciencias F\'\i{}sicas, ECyT-UNSAM and CONICET, Campus Miguelete -- San Mart\'\i{}n, Buenos Aires, Argentina
\item[$^{9}$] Laboratorio Atm\'osfera -- Departamento de Investigaciones en L\'aseres y sus Aplicaciones -- UNIDEF (CITEDEF-CONICET), Argentina
\item[$^{10}$] Observatorio Pierre Auger, Malarg\"ue, Argentina
\item[$^{11}$] Observatorio Pierre Auger and Comisi\'on Nacional de Energ\'\i{}a At\'omica, Malarg\"ue, Argentina
\item[$^{12}$] Universidad Tecnol\'ogica Nacional -- Facultad Regional Buenos Aires, Buenos Aires, Argentina
\item[$^{13}$] University of Adelaide, Adelaide, S.A., Australia
\item[$^{14}$] Universit\'e Libre de Bruxelles (ULB), Brussels, Belgium
\item[$^{15}$] Vrije Universiteit Brussels, Brussels, Belgium
\item[$^{16}$] Centro Brasileiro de Pesquisas Fisicas, Rio de Janeiro, RJ, Brazil
\item[$^{17}$] Centro Federal de Educa\c{c}\~ao Tecnol\'ogica Celso Suckow da Fonseca, Petropolis, Brazil
\item[$^{18}$] Instituto Federal de Educa\c{c}\~ao, Ci\^encia e Tecnologia do Rio de Janeiro (IFRJ), Brazil
\item[$^{19}$] Universidade de S\~ao Paulo, Escola de Engenharia de Lorena, Lorena, SP, Brazil
\item[$^{20}$] Universidade de S\~ao Paulo, Instituto de F\'\i{}sica de S\~ao Carlos, S\~ao Carlos, SP, Brazil
\item[$^{21}$] Universidade de S\~ao Paulo, Instituto de F\'\i{}sica, S\~ao Paulo, SP, Brazil
\item[$^{22}$] Universidade Estadual de Campinas (UNICAMP), IFGW, Campinas, SP, Brazil
\item[$^{23}$] Universidade Estadual de Feira de Santana, Feira de Santana, Brazil
\item[$^{24}$] Universidade Federal de Campina Grande, Centro de Ciencias e Tecnologia, Campina Grande, Brazil
\item[$^{25}$] Universidade Federal do ABC, Santo Andr\'e, SP, Brazil
\item[$^{26}$] Universidade Federal do Paran\'a, Setor Palotina, Palotina, Brazil
\item[$^{27}$] Universidade Federal do Rio de Janeiro, Instituto de F\'\i{}sica, Rio de Janeiro, RJ, Brazil
\item[$^{28}$] Universidad de Medell\'\i{}n, Medell\'\i{}n, Colombia
\item[$^{29}$] Universidad Industrial de Santander, Bucaramanga, Colombia
\item[$^{30}$] Charles University, Faculty of Mathematics and Physics, Institute of Particle and Nuclear Physics, Prague, Czech Republic
\item[$^{31}$] Institute of Physics of the Czech Academy of Sciences, Prague, Czech Republic
\item[$^{32}$] Palacky University, Olomouc, Czech Republic
\item[$^{33}$] CNRS/IN2P3, IJCLab, Universit\'e Paris-Saclay, Orsay, France
\item[$^{34}$] Laboratoire de Physique Nucl\'eaire et de Hautes Energies (LPNHE), Sorbonne Universit\'e, Universit\'e de Paris, CNRS-IN2P3, Paris, France
\item[$^{35}$] Univ.\ Grenoble Alpes, CNRS, Grenoble Institute of Engineering Univ.\ Grenoble Alpes, LPSC-IN2P3, 38000 Grenoble, France
\item[$^{36}$] Universit\'e Paris-Saclay, CNRS/IN2P3, IJCLab, Orsay, France
\item[$^{37}$] Bergische Universit\"at Wuppertal, Department of Physics, Wuppertal, Germany
\item[$^{38}$] Karlsruhe Institute of Technology (KIT), Institute for Experimental Particle Physics, Karlsruhe, Germany
\item[$^{39}$] Karlsruhe Institute of Technology (KIT), Institut f\"ur Prozessdatenverarbeitung und Elektronik, Karlsruhe, Germany
\item[$^{40}$] Karlsruhe Institute of Technology (KIT), Institute for Astroparticle Physics, Karlsruhe, Germany
\item[$^{41}$] RWTH Aachen University, III.\ Physikalisches Institut A, Aachen, Germany
\item[$^{42}$] Universit\"at Hamburg, II.\ Institut f\"ur Theoretische Physik, Hamburg, Germany
\item[$^{43}$] Universit\"at Siegen, Department Physik -- Experimentelle Teilchenphysik, Siegen, Germany
\item[$^{44}$] Gran Sasso Science Institute, L'Aquila, Italy
\item[$^{45}$] INFN Laboratori Nazionali del Gran Sasso, Assergi (L'Aquila), Italy
\item[$^{46}$] INFN, Sezione di Catania, Catania, Italy
\item[$^{47}$] INFN, Sezione di Lecce, Lecce, Italy
\item[$^{48}$] INFN, Sezione di Milano, Milano, Italy
\item[$^{49}$] INFN, Sezione di Napoli, Napoli, Italy
\item[$^{50}$] INFN, Sezione di Roma ``Tor Vergata'', Roma, Italy
\item[$^{51}$] INFN, Sezione di Torino, Torino, Italy
\item[$^{52}$] Istituto di Astrofisica Spaziale e Fisica Cosmica di Palermo (INAF), Palermo, Italy
\item[$^{53}$] Osservatorio Astrofisico di Torino (INAF), Torino, Italy
\item[$^{54}$] Politecnico di Milano, Dipartimento di Scienze e Tecnologie Aerospaziali , Milano, Italy
\item[$^{55}$] Universit\`a del Salento, Dipartimento di Matematica e Fisica ``E.\ De Giorgi'', Lecce, Italy
\item[$^{56}$] Universit\`a dell'Aquila, Dipartimento di Scienze Fisiche e Chimiche, L'Aquila, Italy
\item[$^{57}$] Universit\`a di Catania, Dipartimento di Fisica e Astronomia ``Ettore Majorana``, Catania, Italy
\item[$^{58}$] Universit\`a di Milano, Dipartimento di Fisica, Milano, Italy
\item[$^{59}$] Universit\`a di Napoli ``Federico II'', Dipartimento di Fisica ``Ettore Pancini'', Napoli, Italy
\item[$^{60}$] Universit\`a di Palermo, Dipartimento di Fisica e Chimica ''E.\ Segr\`e'', Palermo, Italy
\item[$^{61}$] Universit\`a di Roma ``Tor Vergata'', Dipartimento di Fisica, Roma, Italy
\item[$^{62}$] Universit\`a Torino, Dipartimento di Fisica, Torino, Italy
\item[$^{63}$] Benem\'erita Universidad Aut\'onoma de Puebla, Puebla, M\'exico
\item[$^{64}$] Unidad Profesional Interdisciplinaria en Ingenier\'\i{}a y Tecnolog\'\i{}as Avanzadas del Instituto Polit\'ecnico Nacional (UPIITA-IPN), M\'exico, D.F., M\'exico
\item[$^{65}$] Universidad Aut\'onoma de Chiapas, Tuxtla Guti\'errez, Chiapas, M\'exico
\item[$^{66}$] Universidad Michoacana de San Nicol\'as de Hidalgo, Morelia, Michoac\'an, M\'exico
\item[$^{67}$] Universidad Nacional Aut\'onoma de M\'exico, M\'exico, D.F., M\'exico
\item[$^{68}$] Institute of Nuclear Physics PAN, Krakow, Poland
\item[$^{69}$] University of \L{}\'od\'z, Faculty of High-Energy Astrophysics,\L{}\'od\'z, Poland
\item[$^{70}$] Laborat\'orio de Instrumenta\c{c}\~ao e F\'\i{}sica Experimental de Part\'\i{}culas -- LIP and Instituto Superior T\'ecnico -- IST, Universidade de Lisboa -- UL, Lisboa, Portugal
\item[$^{71}$] ``Horia Hulubei'' National Institute for Physics and Nuclear Engineering, Bucharest-Magurele, Romania
\item[$^{72}$] Institute of Space Science, Bucharest-Magurele, Romania
\item[$^{73}$] Center for Astrophysics and Cosmology (CAC), University of Nova Gorica, Nova Gorica, Slovenia
\item[$^{74}$] Experimental Particle Physics Department, J.\ Stefan Institute, Ljubljana, Slovenia
\item[$^{75}$] Universidad de Granada and C.A.F.P.E., Granada, Spain
\item[$^{76}$] Instituto Galego de F\'\i{}sica de Altas Enerx\'\i{}as (IGFAE), Universidade de Santiago de Compostela, Santiago de Compostela, Spain
\item[$^{77}$] IMAPP, Radboud University Nijmegen, Nijmegen, The Netherlands
\item[$^{78}$] Nationaal Instituut voor Kernfysica en Hoge Energie Fysica (NIKHEF), Science Park, Amsterdam, The Netherlands
\item[$^{79}$] Stichting Astronomisch Onderzoek in Nederland (ASTRON), Dwingeloo, The Netherlands
\item[$^{80}$] Universiteit van Amsterdam, Faculty of Science, Amsterdam, The Netherlands
\item[$^{81}$] Case Western Reserve University, Cleveland, OH, USA
\item[$^{82}$] Colorado School of Mines, Golden, CO, USA
\item[$^{83}$] Department of Physics and Astronomy, Lehman College, City University of New York, Bronx, NY, USA
\item[$^{84}$] Michigan Technological University, Houghton, MI, USA
\item[$^{85}$] New York University, New York, NY, USA
\item[$^{86}$] University of Chicago, Enrico Fermi Institute, Chicago, IL, USA
\item[$^{87}$] University of Delaware, Department of Physics and Astronomy, Bartol Research Institute, Newark, DE, USA
\item[] -----
\item[$^{a}$] Max-Planck-Institut f\"ur Radioastronomie, Bonn, Germany
\item[$^{b}$] also at Kapteyn Institute, University of Groningen, Groningen, The Netherlands
\item[$^{c}$] School of Physics and Astronomy, University of Leeds, Leeds, United Kingdom
\item[$^{d}$] Fermi National Accelerator Laboratory, Fermilab, Batavia, IL, USA
\item[$^{e}$] Pennsylvania State University, University Park, PA, USA
\item[$^{f}$] Colorado State University, Fort Collins, CO, USA
\item[$^{g}$] Louisiana State University, Baton Rouge, LA, USA
\item[$^{h}$] now at Graduate School of Science, Osaka Metropolitan University, Osaka, Japan
\item[$^{i}$] Institut universitaire de France (IUF), France
\item[$^{j}$] now at Technische Universit\"at Dortmund and Ruhr-Universit\"at Bochum, Dortmund and Bochum, Germany
\end{description}

% created on 2025-06-06
\section*{Acknowledgments}

\begin{sloppypar}
The successful installation, commissioning, and operation of the Pierre
Auger Observatory would not have been possible without the strong
commitment and effort from the technical and administrative staff in
Malarg\"ue. We are very grateful to the following agencies and
organizations for financial support:
\end{sloppypar}

\begin{sloppypar}
Argentina -- Comisi\'on Nacional de Energ\'\i{}a At\'omica; Agencia Nacional de
Promoci\'on Cient\'\i{}fica y Tecnol\'ogica (ANPCyT); Consejo Nacional de
Investigaciones Cient\'\i{}ficas y T\'ecnicas (CONICET); Gobierno de la
Provincia de Mendoza; Municipalidad de Malarg\"ue; NDM Holdings and Valle
Las Le\~nas; in gratitude for their continuing cooperation over land
access; Australia -- the Australian Research Council; Belgium -- Fonds
de la Recherche Scientifique (FNRS); Research Foundation Flanders (FWO),
Marie Curie Action of the European Union Grant No.~101107047; Brazil --
Conselho Nacional de Desenvolvimento Cient\'\i{}fico e Tecnol\'ogico (CNPq);
Financiadora de Estudos e Projetos (FINEP); Funda\c{c}\~ao de Amparo \`a
Pesquisa do Estado de Rio de Janeiro (FAPERJ); S\~ao Paulo Research
Foundation (FAPESP) Grants No.~2019/10151-2, No.~2010/07359-6 and
No.~1999/05404-3; Minist\'erio da Ci\^encia, Tecnologia, Inova\c{c}\~oes e
Comunica\c{c}\~oes (MCTIC); Czech Republic -- GACR 24-13049S, CAS LQ100102401,
MEYS LM2023032, CZ.02.1.01/0.0/0.0/16{\textunderscore}013/0001402,
CZ.02.1.01/0.0/0.0/18{\textunderscore}046/0016010 and
CZ.02.1.01/0.0/0.0/17{\textunderscore}049/0008422 and CZ.02.01.01/00/22{\textunderscore}008/0004632;
France -- Centre de Calcul IN2P3/CNRS; Centre National de la Recherche
Scientifique (CNRS); Conseil R\'egional Ile-de-France; D\'epartement
Physique Nucl\'eaire et Corpusculaire (PNC-IN2P3/CNRS); D\'epartement
Sciences de l'Univers (SDU-INSU/CNRS); Institut Lagrange de Paris (ILP)
Grant No.~LABEX ANR-10-LABX-63 within the Investissements d'Avenir
Programme Grant No.~ANR-11-IDEX-0004-02; Germany -- Bundesministerium
f\"ur Bildung und Forschung (BMBF); Deutsche Forschungsgemeinschaft (DFG);
Finanzministerium Baden-W\"urttemberg; Helmholtz Alliance for
Astroparticle Physics (HAP); Helmholtz-Gemeinschaft Deutscher
Forschungszentren (HGF); Ministerium f\"ur Kultur und Wissenschaft des
Landes Nordrhein-Westfalen; Ministerium f\"ur Wissenschaft, Forschung und
Kunst des Landes Baden-W\"urttemberg; Italy -- Istituto Nazionale di
Fisica Nucleare (INFN); Istituto Nazionale di Astrofisica (INAF);
Ministero dell'Universit\`a e della Ricerca (MUR); CETEMPS Center of
Excellence; Ministero degli Affari Esteri (MAE), ICSC Centro Nazionale
di Ricerca in High Performance Computing, Big Data and Quantum
Computing, funded by European Union NextGenerationEU, reference code
CN{\textunderscore}00000013; M\'exico -- Consejo Nacional de Ciencia y Tecnolog\'\i{}a
(CONACYT) No.~167733; Universidad Nacional Aut\'onoma de M\'exico (UNAM);
PAPIIT DGAPA-UNAM; The Netherlands -- Ministry of Education, Culture and
Science; Netherlands Organisation for Scientific Research (NWO); Dutch
national e-infrastructure with the support of SURF Cooperative; Poland
-- Ministry of Education and Science, grants No.~DIR/WK/2018/11 and
2022/WK/12; National Science Centre, grants No.~2016/22/M/ST9/00198,
2016/23/B/ST9/01635, 2020/39/B/ST9/01398, and 2022/45/B/ST9/02163;
Portugal -- Portuguese national funds and FEDER funds within Programa
Operacional Factores de Competitividade through Funda\c{c}\~ao para a Ci\^encia
e a Tecnologia (COMPETE); Romania -- Ministry of Research, Innovation
and Digitization, CNCS-UEFISCDI, contract no.~30N/2023 under Romanian
National Core Program LAPLAS VII, grant no.~PN 23 21 01 02 and project
number PN-III-P1-1.1-TE-2021-0924/TE57/2022, within PNCDI III; Slovenia
-- Slovenian Research Agency, grants P1-0031, P1-0385, I0-0033, N1-0111;
Spain -- Ministerio de Ciencia e Innovaci\'on/Agencia Estatal de
Investigaci\'on (PID2019-105544GB-I00, PID2022-140510NB-I00 and
RYC2019-027017-I), Xunta de Galicia (CIGUS Network of Research Centers,
Consolidaci\'on 2021 GRC GI-2033, ED431C-2021/22 and ED431F-2022/15),
Junta de Andaluc\'\i{}a (SOMM17/6104/UGR and P18-FR-4314), and the European
Union (Marie Sklodowska-Curie 101065027 and ERDF); USA -- Department of
Energy, Contracts No.~DE-AC02-07CH11359, No.~DE-FR02-04ER41300,
No.~DE-FG02-99ER41107 and No.~DE-SC0011689; National Science Foundation,
Grant No.~0450696, and NSF-2013199; The Grainger Foundation; Marie
Curie-IRSES/EPLANET; European Particle Physics Latin American Network;
and UNESCO.
The authors gratefully acknowledge the
computing time provided on the high-performance computer HoreKa by the National High-Performance Computing
Center at KIT (NHR@KIT). This center is jointly supported by the Federal Ministry of Education and Research and
the Ministry of Science, Research and the Arts of Baden-Wüurttemberg, as part of the National High-Performance
Computing (NHR) joint funding program. HoreKa is partly funded by the German Research Foundation.
\end{sloppypar}

}

\end{document}